\documentclass[eqsecnum,showpacs,showkeys,nofootinbib,aps]{revtex4}

\usepackage{epsfig}

\renewcommand{\theequation}{\arabic{equation}}
\def\be{\begin{equation}}
\def\ee{\end{equation}}
\def\bea{\begin{eqnarray}}
\def\eea{\end{eqnarray}}
\def\nn{\nonumber}

\def\na{\nabla}

\begin{document}

\title{Local free-fall temperatures of charged BTZ black holes in massive gravity}
\author{Soon-Tae Hong}
\email{galaxy.mass@gmail.com}
\affiliation{Center for Quantum Spacetime and
\\ Department of Physics, Sogang University, Seoul 04107, Korea}
\author{Yong-Wan Kim}
\email{ywkim65@gmail.com}
 \affiliation{Department of Physics and
 \\ Research Institute of Physics and Chemistry, Chonbuk National University, Jeonju 54896, Korea}
\author{Young-Jai Park}
\email{yjpark@sogang.ac.kr}
\affiliation{Center for Quantum Spacetime and
\\ Department of Physics, Sogang University, Seoul 04107, Korea }
\date{\today}

\begin{abstract}
We obtain a (3+3)-dimensional global flat embedding of the
generalized (2+1) charged Ba\~nados-Teitelboim-Zanelli black holes
in massive gravity. We also study the local free-fall temperatures
for freely falling observers starting from rest and investigate
the effect of the charge and graviton mass in free-fall
temperatures.

\end{abstract}

\maketitle

\section{introduction}
\setcounter{equation}{0}
\renewcommand{\theequation}{\arabic{section}.\arabic{equation}}

As discovered by Hawking \cite{Hawking:1974sw}, an observer
located at asymptotic infinity sees the Hawking temperature $T_H$
of a black hole that emits characteristic thermal radiation. On
the other hand, a fiducial observer at a finite distance from a
black hole sees a local temperature described by the Tolman
temperature \cite{Tolman:1987}
 \be
 T_{\rm FID} = \frac{T_H}{\sqrt{g_{\mu\nu}\xi^\mu\xi^\nu}},
 \ee
where $\xi^\mu$ is a timelike Killing vector. Later, Unruh
\cite{Unruh:1976db} showed that a uniformly accelerating observer
in a flat spacetime, with a proper acceleration $a$, detects
thermal radiation at the Unruh temperature
 \be
 T_U = \frac{a}{2\pi}.
 \ee
These two effects are related; {\it i.e.}, the Hawking effect for
a fiducial observer in a black hole spacetime can be considered as
the Unruh effect for a uniformly accelerated observer in a
higher-dimensional global embedding Minkowski spacetime (GEMS).
These ideas and their corresponding GEMSs are studied through the
analysis of de Sitter \cite{Narnhofer:1996zk} and anti-de Sitter
(AdS) spacetimes \cite{Deser:1997ri,Deser:1998bb}. Furthermore,
Deser and Levin \cite{Deser:1998xb} have shown that the GEMS
approach provides a unified derivation of temperature for
uncharged Ba\~nados-Teitelboim-Zanelli (BTZ)
\cite{Banados:1992wn,Banados:1992gq}, Schwarzschild-AdS, and
Reissner-Nordstr\"om (RN) spacetimes. After these works, we have
constructed GEMSs for the charged BTZ~\cite{Hong:2000kn} and
RN-AdS~\cite{Kim:2000ct} spacetimes according to this approach.
Since then, there have been many works on a variety of curved
spacetimes~\cite{Hong:2003xz,Chen:2004qw,Santos:2004ws,
Banerjee:2010ma,Cai:2010bv,Majhi:2011yi,Hu:2011yx,hong2000,hong2001,
hong2005,hong2003,hong2004,hong2006,paston2014}. Furthermore,
several years ago, Brynjolfsson and Thorlacius
\cite{Brynjolfsson:2008uc} used the GEMS approach to define a
local temperature for a freely falling observer outside
Schwarzschild(-AdS) and RN spacetimes, showing that freely falling
temperatures remain finite at event horizons while they approach
the Hawking temperatures at asymptotic infinities. Here, a freely
falling local temperature is defined at special turning points of
radial geodesics where a freely falling observer is momentarily at
rest with respect to a black hole. After the work, we have
extended the results to RN-AdS \cite{Kim:2009ha},
Gibbons-Maeda-Garfinkle-Horowitz-Strominger black holes
\cite{Kim:2013wpa}, a modified Schwarzschild black hole in rainbow
spacetime \cite{Kim:2015wwa}, and a Schwarzschild-Tangherlini-AdS
black hole~\cite{hong2015}. However, up to now, all these studies
of finding freely falling temperatures have been mainly restricted
to massless graviton cases. By the way, it is known that massive
gravitons in general relativity~\cite{hinter2012,vegh2013} can be
introduced by various channels, one of which is breaking the
Lorentz symmetry of the system~\cite{Chamseddine:2012gh}. The
modification to the behavior of a black hole by including a
graviton mass has also been considered in the extended phase space
in order to study the phase transition of black
holes~\cite{Hendi:2015eca}. Moreover, the graviton mass terms have
been exploited to investigate many interesting models such as, for
instance, Gauss-Bonnet massive gravity~\cite{Hendi:2015pda}.  It
has been noticed that the massive gravitons can yield interesting
modification of black hole thermodynamics.

On the other hand, since the pioneering work in 1992, the
(2+1)-dimensional BTZ black hole in massless
gravity~\cite{Banados:1992wn,Banados:1992gq} has become a useful
model for realistic black hole physics~\cite{Carlip:1995qv}.
Moreover, significant interest in this model has recently
increased with the discovery that the thermodynamics of
higher-dimensional black holes can often be interpreted in terms
of the BTZ solution~\cite{Hyun:1997jv}. It is therefore
interesting to study the geometry of (2+1)-dimensional black holes
and their thermodynamics through further investigation. Moreover,
it is possible to construct a BTZ black hole in massive
gravity~\cite{hendi2016,Hendi:2016hbe,zaz2017,prasia2017,chougule2018}.
In fact, an asymptotically AdS charged BTZ black hole has been
constructed in a massive theory of gravity, and various different
aspects of such a solution have been studied. In these works, they
have considered three-dimensional massive gravity with an Abelian
U(1) gauge field and negative cosmological constant of which the
action is of the form
 \be S=-\frac{1}{16}\int d^{3}x\sqrt{g}\left[{\cal R}
   -2\Lambda+{\cal L}({\cal F})+\tilde{M}^{2}\sum_{i=1}^{4}c_{i}{\cal U}_{i}(g,f)\right],
   \label{massivebtz}
 \ee
where ${\cal R}$ is the scalar curvature, $\Lambda(=-1/l^{2})$ is
the cosmological constant, ${\cal L}({\cal F})$ is the Lagrangian
for the vector gauge field, $\tilde{M}$ is the massive parameter,
and $f$ is the reference metric. ${\cal F}(=F_{\mu\nu}F^{\mu\nu})$
is the Maxwell invariant in which
$F_{\mu\nu}(=\partial_{\mu}A_{\nu}-\partial_{\nu}A_{\mu})$ is the
Faraday tensor and $A_{\mu}$ is the U(1) gauge potential. $c_{i}$
are the constants for massive gravity, and ${\cal U}_{i}$ are the
symmetric polynomials of eigenvalues. Here, they take an ansatz
that ${\cal U}_{1}=c/r$, ${\cal U}_{2}={\cal U}_{3}={\cal
U}_{4}=0$, where $c$ is a positive constant.

In this paper, we will generalize the Unruh, Hawking, and freely
falling temperatures of the charged BTZ black hole in the massless
case\footnote{In this work, we will call it massless when
$\tilde{M}$ is zero .} to those in the massive gravity
(\ref{massivebtz}) with the ansatz in terms of the GEMS approach.
In Sec. II, we will briefly summarize the GEMS embedding of the
charged BTZ black hole in the massless gravity~\cite{Hong:2000kn}
and then newly obtain desired temperatures of the black holes as
measured by freely falling observers. In Secs. III and IV, we will
derive the GEMS embeddings of the uncharged and charged BTZ black
holes in the massive gravity and then evaluate local temperatures
of the black holes as measured by freely falling observers,
respectively. In particular, in Sec. IV, we discuss the effect of
charge and massive gravitons on the Hawking temperature in the
charged BTZ black hole in the massive gravity. Finally, our
conclusions are drawn in Sec. V.

\section{Charged BTZ black hole in massless gravity}
\setcounter{equation}{0}
\renewcommand{\theequation}{\arabic{section}.\arabic{equation}}

\subsection{GEMS of charged BTZ black hole}

We consider the (2+1)-dimensional charged BTZ black hole in the
massless gravity described by the 3-metric
 \be
 ds^2=N^2dt^2-N^{-2}dr^{2}-r^{2}d{\phi}^{2} \label{3metric}
 \ee
with the lapse function
 \be
 N^{2}=-m+\frac{r^{2}}{l^{2}}-2q^{2}\ln \frac{r}{l} \label{cbtzmetric},
 \ee
where $m=8M$ and $q=2Q$ with $M$ and $Q$ being the mass and
electric charge of the BTZ black hole, respectively. Now, the mass
$m$ can be written in terms of the event horizon $r_{H}$ as \be
m=\frac{r_{H}^{2}}{l^{2}}-2q^{2}\ln \frac{r_{H}}{l}, \ee and the
Hawking-Bekenstein horizon surface gravity is given by~\cite{wald}
 \be
 k_H = -\frac{1}{2}(\na^{\mu}\xi^{\nu})(\na_{\mu}\xi_{\nu})|_{r\rightarrow r_H}
     =\frac{r_{H}}{l^{2}}-\frac{q^{2}}{r_{H}}.
\label{kh0}
 \ee

Then, according to the GEMS approach, a (3+3)-dimensional AdS GEMS
 \be
ds^{2}=(dz^{0})^2-(dz^{1})^2
  -(dz^{2})^2+(dz^{3})^{2}-(dz^{4})^{2}+(dz^{5})^{2}
\ee
is given by the coordinate transformations for $r\geq r_{H}$ \cite{Hong:2000kn}
 \bea
z^{0}&=&k_{H}^{-1}\left(\frac{r^{2}-r_{H}^{2}}{l^{2}}-2q^{2}
\ln\frac{r}{r_{H}}\right)^{1/2}\sinh k_{H}t, \nonumber \\
z^{1}&=&k_{H}^{-1}\left(\frac{r^{2}-r_{H}^{2}}{l^{2}}-2q^{2}
\ln\frac{r}{r_{H}}\right)^{1/2}\cosh k_{H}t, \nonumber \\
z^{2}&=&\frac{l}{r_{H}}r\sinh \frac{r_{H}}{l}\phi, \nonumber \\
z^{3}&=&\frac{l}{r_{H}}r\cosh \frac{r_{H}}{l}\phi, \nonumber \\
z^{4}&=&k_{H}^{-1}\int dr \frac{q^{2}l[r^{2}+r_{H}^{2}+2r^{2}g(r)]^{1/2}}
{r_{H}^{2}r\left[1-\frac{q^{2}l^{2}}{r_{H}^{2}}g(r)\right]^{1/2}}, \nonumber \\
z^{5}&=&k_{H}^{-1}\int dr \frac{q\left[2r_{H}^{2}+\frac{r_{H}^{4}+q^{4}l^{4}}{r_{H}^{2}}g(r)\right]^{1/2}}
{r_{H}^{2}\left[1-\frac{q^{2}l^{2}}{r_{H}^{2}}g(r)\right]^{1/2}},
 \label{gemsbtz0}
 \eea
where \be g(r)=\frac{2r_{H}^{2}}{r^{2}-r_{H}^{2}}\ln
\frac{r}{r_{H}}. \label{gr0} \ee Here one notes that, due to
l'H\^opital's rule, $g(r)$ approaches unity as $r$ goes to
$r=r_{H}$.

For the trajectories, which follow the Killing vector
$\xi=\partial_{t}$, we can obtain a constant 3-acceleration
 \be
 a_{3}=\frac{\frac{r}{l^{2}}-\frac{q^{2}}{r}}{\left(\frac{r^{2}-r_{H}^{2}}{l^{2}}-2q^{2}\ln \frac{r}{r_{H}}\right)^{1/2}}.
\label{a30}
 \ee
In static detectors ($\phi$, $r=$ const) described by a fixed
point in the ($z^{2}$, $z^{3}$, $z^{4}$, $z^{5}$) plane, an
observer, who is uniformly accelerated in the (3+3)-dimensional
flat spacetime, follows a hyperbolic trajectory in
($z^{0}$,$z^{1}$) described by a proper acceleration $a_{6}$ as
follows:
 \be
 a^{-2}_{6}=(z^1)^2-(z^0)^2=\frac{l^{2}\left(r^{2}-r_{H}^{2}-2q^{2}l^{2}\ln \frac{r}{r_{H}}\right)}
{(r_{H}-\frac{q^{2}l^{2}}{r_{H}})^{2}}.\label{sbtz-accel0}
 \ee
Here, we have the relation with a constant acceleration $a_{3}$,
 \be
 a_{6}^{2}-a_{3}^{2}=-\frac{1}{l^{2}}
   +\frac{\frac{q^{4}l^{2}}{r^{2}r_{H}^{2}}-\frac{q^{2}}{r_{H}^{2}}g(r)}
{1-\frac{q^{2}l^{2}}{r_{H}^{2}}g(r)}. \label{a3a20} \ee One notes
that, in the limit of $q=0$, $a_{6}^{2}-a_{3}^{2}$ becomes
$-1/l^{2}$ as expected~\cite{ Deser:1998xb,Hong:2000kn}.

As was shown by Unruh~\cite{Unruh:1976db}, the Unruh temperature
for a uniformly accelerated observer in the (3+3)-dimensional flat
spacetime can be read as
 \be
 T_U=\frac{a_{6}}{2\pi},
\label{ths0}
 \ee
so we can obtain
 \be
 T_U=\frac{r_{H}-\frac{q^{2}l^{2}}{r_{H}}}
{2\pi l\left(r^{2}-r_{H}^{2}-2q^{2}l^{2}\ln \frac{r}{r_{H}}\right)^{1/2}}.
\label{tu0}
 \ee
This is exactly the same with the local temperature measured by a
fiducial observer staying at a finite distance from the black
hole, the so-called fiducial temperature,
 \be
 T_{\rm FID}=\frac{T_H}{\sqrt{g_{00}}},
\label{tfid0}
 \ee
where the Hawking temperature $T_{H}$ is measured by an asymptotic
observer,
 \be
 T_H=\frac{1}{2\pi}\left(\frac{r_{H}}{l^{2}}-\frac{q^{2}}{r_{H}}\right).
\label{th0}
 \ee
Next, by introducing dimensionless variables
 \be
 x=\frac{r_{H}}{r},~~~a=\frac{l}{r_{H}},~~~b=q^{2},
 \label{dimensionless0}
 \ee
we can rewrite the Hawking temperature in Eq. (\ref{th0}) as
follows:
 \be
 T_H\cdot r_{H}=\frac{1}{2\pi}
   \left(\frac{1}{a^{2}}-b\right).
  \label{thdimensionless}
 \ee

\subsection{Free-fall temperature of charged BTZ black hole}

Now, we assume that an observer at rest is freely falling from the
radial position $r=r_{0}$ at
$\tau=0$~\cite{Brynjolfsson:2008uc,Kim:2009ha,Kim:2013wpa,Kim:2015wwa,hong2015}.
The equations of motion for the orbit of the observer are given as
 \bea
 \frac{dt}{d\tau}&=&\frac{N(r_{0})}{N^{2}(r)},\nn\\
 \frac{dr}{d\tau}&=&-\left[N^{2}(r_{0})-N^{2}(r)\right]^{1/2}.
\label{eomr0}
 \eea
Exploiting Eqs. (\ref{gemsbtz0}) and (\ref{eomr0}), we obtain the
freely falling acceleration $\bar{a}_{6}$ in the (3+3)-dimensional
GEMS embedded spacetime
 \be
 \bar{a}_{6}=\frac{1}{l}\left[\frac{\left(1+\frac{q^{2}l^{2}}{rr_{H}}\right)
\left(1-\frac{q^{2}l^{2}}{rr_{H}}\right)}
{1-\frac{q^{2}l^{2}}{r_{H}^{2}}g(r)}\right]^{1/2},
\label{a4btz0}
 \ee
which gives us the temperature measured by the freely falling observer
at rest (FFAR)
 \be
  T_{\rm FFAR}=\frac{\bar{a}_{6}}{2\pi}=\frac{1}{2\pi l}\left[\frac{\left(1+\frac{q^{2}l^{2}}{rr_{H}}\right)
\left(1-\frac{q^{2}l^{2}}{rr_{H}}\right)}
{1-\frac{q^{2}l^{2}}{r_{H}^{2}}g(r)}\right]^{1/2}.
 \label{tffarbtzdef0massless}
 \ee
It is appropriate to comment that in the limit of $q=0$, free-fall
temperature for an uncharged BTZ black hole seen by the freely
falling observer is reduced to
 \be T_{\rm FFAR}\cdot r_{H}=\frac{r_{H}}{2\pi l}=\frac{1}{2\pi a}.
 \label{uchargedtff}
 \ee

Making use of the dimensionless variables introduced in Eq.
(\ref{dimensionless0}), we can rewrite the free-fall temperature
in Eq. (\ref{tffarbtzdef0massless}) as
 \be
 T_{\rm FFAR}\cdot r_{H}=\frac{1}{2\pi a}
   \left[\frac{(1+a^{2}bx)\left(1-a^{2}bx\right)}
  {1+\frac{2a^{2}bx^{2}}{1-x^{2}}\ln x}\right]^{1/2},
  \label{tffrhdimensionless}
 \ee
where the relevant range of $x$ is given by $0\le x\le 1$.

\begin{figure*}[t!]
   \centering
   \includegraphics{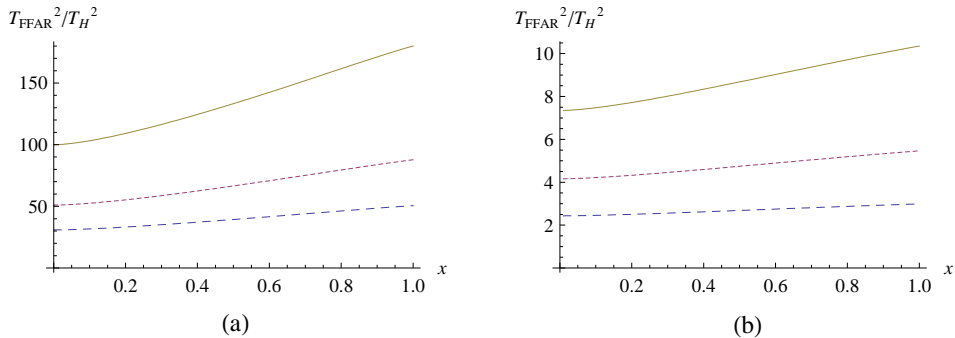}
\caption{Free-fall temperature for the charged BTZ black hole in
the massless gravity: (a) for a fixed $a=2$ with
$b=0.16,~0.18,~0.20$; (b) for a fixed $b=0.16$ with
$a=1.2,~1.4,~1.6$ from bottom to top, respectively. }
 \label{fig1}
\end{figure*}

In the limit of $x=0$, we obtain \be T_{\rm FFAR}\cdot
r_{H}~(x=0)=\frac{1}{2\pi a}, \label{tffarx0massless} \ee where we
have used the identity $\lim_{x\rightarrow 0}x^{2}\ln x =0$. Here,
we note that the above value in Eq. (\ref{tffarx0massless}) is
independent of the parameters $b$. This means that, as
$r\rightarrow\infty$, the charge of the charged BTZ black hole
does not affect the free-fall temperature $T_{\rm FFAR}$. In other
words, the free-fall temperature $T_{\rm FFAR}$ as
$r\rightarrow\infty$ in Eq. (\ref{tffarx0massless}) is the same as
that of the uncharged BTZ black hole in Eq. (\ref{uchargedtff}).

At $x=1$, namely, at the event horizon $r=r_{H}$, we end up with
\be T_{\rm FFAR}\cdot r_{H}~(x=1)=\frac{1}{2\pi
a}\left(1+a^{2}b\right)^{1/2}, \label{tffarrhmassless} \ee where
we have exploited the relation $\lim_{x\rightarrow 1}\frac{\ln
x}{1-x^{2}}=-\frac{1}{2}$. The above result in Eq.
(\ref{tffarrhmassless}) shows that there exists no singularity of
$T_{\rm FFAR}$ at $x=1$. Moreover, we note that \be T_{\rm
FFAR}~\left(x=1\right)>T_{\rm FFAR}~(x=0),
\label{tffarrhcancelmassless} \ee which means that the free-fall
temperature at the event horizon $r=r_{H}$ is greater than that at
the infinity, as expected. These are summarized in Fig. \ref{fig1}
by plotting $T^2_{\rm FFAR}$ in units of $T^2_H$. Here, Fig.
\ref{fig1}(a) was drawn by changing the charge $b$, while Fig.
\ref{fig1}(b) was drawn by changing the cosmological constant $a$.

\section{Uncharged BTZ black hole in massive gravity}
\setcounter{equation}{0}
\renewcommand{\theequation}{\arabic{section}.\arabic{equation}}

\subsection{GEMS of uncharged BTZ black hole in massive gravity}

Now, let us consider a (2+1)-dimensional uncharged BTZ black hole
in the massive gravity described by the 3-metric in Eq.
(\ref{3metric}) with the lapse function
 \be
 N^{2}=-m+\frac{r^{2}}{l^{2}}+2Rr.
 \ee
Here, the notation $R$ related to the mass term in Eq.
(\ref{massivebtz}) is given by
 \be
 R=\frac{1}{2}\tilde{M}^{2}cc_{1}. \label{rmcc}
  \ee
Now, the mass $m$ of the uncharge BTZ black hole in the massive
gravity can be given in terms of the event horizon $r_{H}$ as \be
m=\frac{r_{H}^{2}}{l^{2}}+2Rr_{H}, \ee and the Hawking-Bekenstein
horizon surface gravity is of the form
 \be
 k_H =\frac{r_{H}}{l^{2}}+R.
\label{kh}
 \ee

Exploiting the GEMS approach, we obtain a (3+2)-dimensional AdS
GEMS,
 \be
ds^{2}=(dz^{0})^2-(dz^{1})^2
  -(dz^{2})^2+(dz^{3})^{2}-(dz^{4})^{2},
\ee given by the coordinate transformations for $r\geq r_{H}$ and
$R>0$ as
 \bea
z^{0}&=&k_{H}^{-1}\left[\frac{r^{2}-r_{H}^{2}}{l^{2}}+2R(r-r_{H})\right]^{1/2}\sinh k_{H}t, \nonumber \\
z^{1}&=&k_{H}^{-1}\left[\frac{r^{2}-r_{H}^{2}}{l^{2}}+2R(r-r_{H})\right]^{1/2}\cosh k_{H}t, \nonumber \\
z^{2}&=&\frac{l}{r_{H}}r\sinh \frac{r_{H}}{l}\phi, \nonumber \\
z^{3}&=&\frac{l}{r_{H}}r\cosh \frac{r_{H}}{l}\phi, \nonumber \\
z^{4}&=&k_{H}^{-1}\int dr \frac{\left[R^{2}l^{2}r_{H}^{2}r^{2}\left(1+\frac{4r_{H}}{r+r_{H}}\right)+Rl^{4}f(r)\right]^{1/2}}
{r_{H}^{2}r\left[1+\frac{2Rl^{2}}{r+r_{H}}\right]^{1/2}}, \nonumber \\
 \label{gemsbtz}
 \eea
where
\bea
f&=&\frac{2r_{H}^{3}r^{2}}{l^{4}}+\frac{2r^{2}R^{2}r_{H}^{2}}{r+r_{H}}.\nonumber\\
\label{frrh}
\eea

For the trajectories, which follow the Killing vector
$\xi=\partial_{t}$, we can obtain a constant 3-acceleration,
 \be
 a_{3}=\frac{\frac{r}{l^{2}}+R}{\left[\frac{r^{2}-r_{H}^{2}}{l^{2}}+2R(r-r_{H})\right]^{1/2}}.
 \ee
In static detectors ($\phi$, $r=$ const) described by a fixed
point in the ($z^{2}$, $z^{3}$, $z^{4}$) plane, an observer, who
is uniformly accelerated in the (3+2)-dimensional flat spacetime,
follows a hyperbolic trajectory in ($z^{0}$,$z^{1}$) described by
a proper acceleration $a_{5}$ as follows:
 \be\label{sbtz-accel}
 a^{-2}_{5}=(z^1)^2-(z^0)^2=\frac{l^{2}\left[r^{2}-r_{H}^{2}+2Rl^{2}(r-r_{H})\right]}
{(r_{H}+Rl^{2})^{2}}.
 \ee
Here, we have the relation with a constant acceleration $a_{3}$,
 \be
 a_{5}^{2}-a_{3}^{2}=-\frac{1}{l^{2}},
 \label{relnucmbtz}
 \ee
which is the same as the result for the uncharged BTZ black hole
in the massless gravity~\cite{ Deser:1998xb,Hong:2000kn}.

Exploiting the relation in Eq. (\ref{ths0}), we arrive at the
Unruh temperature for a uniformly accelerated observer in the
(3+2)-dimensional flat spacetime:
 \be
 T_U=\frac{r_{H}+Rl^{2}}
{2\pi l\left[r^{2}-r_{H}^{2}+2Rl^{2}(r-r_{H})\right]^{1/2}}.
 \ee
This is exactly the same with the fiducial temperature $T_{\rm
FID}$ in Eq. (\ref{tfid0}) with $T_{H}$ being the Hawking
temperature measured by an asymptotic observer,
 \be
 T_H=\frac{1}{2\pi}\left(\frac{r_{H}}{l^{2}}+R\right).
 \label{hawkingthr}
 \ee
As a result, one can say that the Hawking effect for a fiducial
observer in the black hole spacetime is equal to the Unruh effect
for a uniformly accelerated observer in a higher-dimensional flat
spacetime. Next, introducing a new additional dimensionless
variable, \be d=Rr_{H}, \label{dimensionless} \ee together with
the other dimensionless variables in Eq. (\ref{dimensionless0}),
we rewrite the Hawking temperature in Eq. (\ref{hawkingthr}) as
follows: \be T_H\cdot
r_{H}=\frac{1}{2\pi}\left(\frac{1}{a^{2}}+d\right). \ee

\subsection{Free-fall temperature of uncharged BTZ black hole in massive gravity}

\begin{figure*}[t!]
 \centering
 \includegraphics{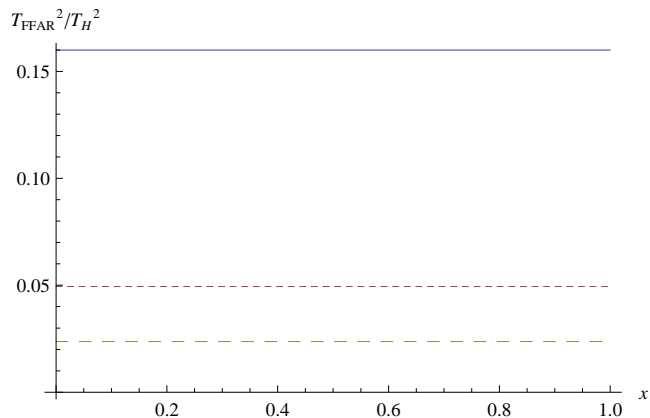}
 \caption{Free-fall temperature for the uncharged BTZ black hole in
 the massive gravity for a fixed $a=2$ with $d=1,~2,~3$ from top to bottom.}
 \label{fig2}
\end{figure*}

Now, we assume that an observer at rest is freely falling from the
radial position $r=r_{0}$ at
$\tau=0$~\cite{Brynjolfsson:2008uc,Kim:2009ha,Kim:2013wpa,Kim:2015wwa,hong2015}.
The equations of motion for the orbit of the observer are given by
Eq. (\ref{eomr0}). Exploiting Eqs. (\ref{gemsbtz}) and
(\ref{eomr0}), we obtain the freely falling acceleration
$\bar{a}_{5}$ in the (3+2)-dimensional GEMS embedded spacetime
 \be
 \bar{a}_{5}=\frac{1}{l},\label{a4btz}
 \ee
which gives us the temperature measured by the freely falling
observer at rest as follows:
 \be
 T_{\rm FFAR}\cdot r_{H}=\frac{1}{2\pi a}.
 \label{tffarbtzdef}
 \ee
This is exactly the same as the result for the uncharged BTZ black
hole in the massless gravity. The squared free-fall temperature
$T^2_{\rm FFAR}$ is depicted in Fig. \ref{fig2} in units of
$T^2_H$.

\section{Charged BTZ black hole in massive gravity}
\setcounter{equation}{0}
\renewcommand{\theequation}{\arabic{section}.\arabic{equation}}

\subsection{GEMS of charged BTZ black hole in massive gravity}

\begin{figure*}[t!]
   \centering
   \includegraphics[scale=0.8]{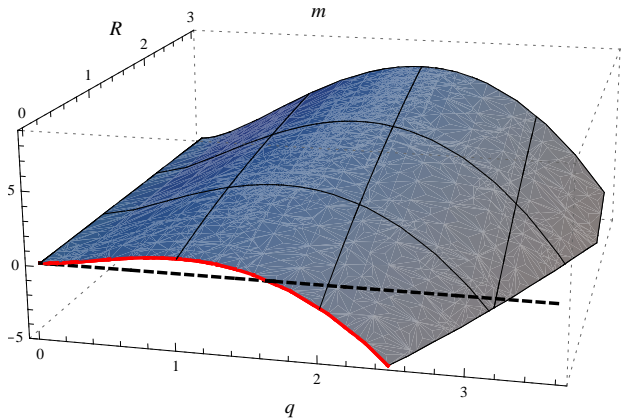}
\caption{Upper region of the surface in the mass-charge relation
in which charged BTZ black holes in massive gravity can exist.}
 \label{fig3}
\end{figure*}

A (2+1)-dimensional charged BTZ black hole in massive gravity is
described by the 3-metric in Eq. (\ref{3metric}) with the lapse
function \cite{hendi2016,chougule2018}
 \be
 N^{2}=-m+\frac{r^{2}}{l^{2}}-2q^{2}\ln \frac{r}{l}+2Rr,
 \label{massivecbtz}
 \ee
where $R$ is given by Eq. (\ref{rmcc}). First of all, it is
appropriate to comment on the metric function (\ref{massivecbtz}),
which goes to positive infinities if $r\rightarrow 0$ and
$r\rightarrow\infty$ so that there is a minimum at
 \be
 r_{\rm min}=\frac{-Rl^2+l\sqrt{4q^2+R^2l^2}}{2}.
 \ee
Thus, the metric function has the value of
 \be
 N^{2}(r)|_{r\rightarrow r_{\rm min}}
        =-m+q^2-\frac{1}{2}R^2l^2+\frac{1}{2}Rl\sqrt{4q^2+R^2l^2}
         -2q^2\ln\left(\frac{-Rl+\sqrt{4q^2+R^2l^2}}{2}\right).
 \label{mcbtzrmin}
 \ee
One can easily see that when $N^{2}(r_{\rm min})<0$ there are two
roots of $r_+$ and $r_-$ and when $N^{2}(r_{\rm min})=0$ the two
roots coincide and one has an extreme black hole. This is depicted
in Fig. {\ref{fig3}}, in which black holes with two horizons exist
over the ($q,R$) surface, an extremal black hole exists at the
($q,R$) surface, and black holes cannot exist below the ($q,R$)
surface. On the figure, the red curve is drawn for $R=0$, in which
case one has some difficulties \cite{Martinez:1999qi} such as a
logarithmic divergent boundary term at $r\rightarrow\infty$ and a
cosmic censorship problem due to having arbitrarily negative
values of $m$. Reference \cite{Cadoni:2009bn} has studied how to
circumvent the problems. For a similar reason, the mass function
in Eq. (\ref{mcbtzrmin}) would have the same problems, which may
be addressed elsewhere. Here, we will obtain the GEMS embedding
only for $r>r_+$ and nonextremal cases.

Now, the mass $m$ in Eq. (\ref{massivecbtz}) can be given in terms
of the event horizon $r_{H}$ as
 \be
 m=\frac{r_{H}^{2}}{l^{2}}-2q^{2}\ln \frac{r_{H}}{l}+2Rr_{H},
 \ee
and the Hawking-Bekenstein horizon surface gravity is of the form
 \be
 k_H =\frac{r_{H}}{l^{2}}-\frac{q^{2}}{r_{H}}+R.
\label{kh4}
 \ee

Exploiting the GEMS approach, we obtain a (3+3)-dimensional AdS
GEMS,
 \be
ds^{2}=(dz^{0})^2-(dz^{1})^2
  -(dz^{2})^2+(dz^{3})^{2}-(dz^{4})^{2}+(dz^{5})^{2},
\ee given by the coordinate transformations for $r\geq r_{H}$ as
 \bea
z^{0}&=&k_{H}^{-1}\left[\frac{r^{2}-r_{H}^{2}}{l^{2}}-2q^{2}
\ln\frac{r}{r_{H}}+2R(r-r_{H})\right]^{1/2}\sinh k_{H}t, \nonumber \\
z^{1}&=&k_{H}^{-1}\left[\frac{r^{2}-r_{H}^{2}}{l^{2}}-2q^{2}
\ln\frac{r}{r_{H}}+2R(r-r_{H})\right]^{1/2}\cosh k_{H}t, \nonumber \\
z^{2}&=&\frac{l}{r_{H}}r\sinh \frac{r_{H}}{l}\phi, \nonumber \\
z^{3}&=&\frac{l}{r_{H}}r\cosh \frac{r_{H}}{l}\phi, \nonumber \\
z^{4}&=&k_{H}^{-1}\int dr \frac{\left[q^{4}l^{2}(r^{2}+r_{H}^{2}+2r^{2}g(r))+R^{2}l^{2}r_{H}^{2}r^{2}\left(1+\frac{4r_{H}}{r+r_{H}}\right)+Rl^{4}f_{1}(r)\right]^{1/2}}
{r_{H}^{2}r\left[1-\frac{q^{2}l^{2}}{r_{H}^{2}}g(r)+\frac{2Rl^{2}}{r+r_{H}}\right]^{1/2}}, \nonumber \\
z^{5}&=&k_{H}^{-1}\int dr \frac{\left[q^{2}r^{2}\left(2r_{H}^{2}+\frac{r_{H}^{4}+q^{4}l^{4}}{r_{H}^{2}}g(r)\right)
+R^{2}q^{2}l^{4}r^{2}\left(g(r)+\frac{4r_{H}}{r+r_{H}}\right)+Rl^{4}f_{2}(r)\right]^{1/2}}
{r_{H}^{2}r\left[1-\frac{q^{2}l^{2}}{r_{H}^{2}}g(r)+\frac{2Rl^{2}}{r+r_{H}}\right]^{1/2}},
 \label{gemsbtz4}
 \eea
where
\bea
f_{1}&=&\frac{2r_{H}^{3}r^{2}}{l^{4}}+\frac{2q^{4}r^{2}}{r_{H}}g(r)+\frac{2r^{2}}{r+r_{H}}(q^{4}+R^{2}r_{H}^{2}),\nonumber\\
f_{2}&=&\frac{2q^{2}r_{H}r^{2}}{l^{2}}[1+g(r)]+\frac{2q^{2}r_{H}^{2}r(2r+r_{H})}{l^{2}(r+r_{H})},
\label{frrh1} \eea and $g(r)$ is given by Eq. (\ref{gr0}). In the
limit of $q=0$, the coordinate transformations in Eq.
(\ref{gemsbtz4}) are reduced to those in Eq. (\ref{gemsbtz}) with
$f_{1}\rightarrow f$ and $f_{2}\rightarrow 0$. Here, note that the
dimensionality (3+3) becomes (3+2) since $z^{5}$ disappears in
this limit.

For the trajectories, which follow the Killing vector
$\xi=\partial_{t}$, we can obtain a constant 3-acceleration:
 \be
 a_{3}=\frac{\frac{r}{l^{2}}-\frac{q^{2}}{r}+R}{\left[\frac{r^{2}-r_{H}^{2}}{l^{2}}-2q^{2}\ln \frac{r}{r_{H}}+2R(r-r_{H})\right]^{1/2}}.
 \ee
In static detectors ($\phi$, $r=$ const) described by a fixed
point in the ($z^{2}$, $z^{3}$, $z^{4}$, $z^{5}$) plane, an
observer, who is uniformly accelerated in the (3+3)-dimensional
flat spacetime, follows a hyperbolic trajectory in
($z^{0}$,$z^{1}$) described by a proper acceleration $a_{6}$ as
follows:
 \be\label{sbtz-accel4}
 a^{-2}_{6}=(z^1)^2-(z^0)^2=\frac{l^{2}\left[r^{2}-r_{H}^{2}-2q^{2}l^{2}\ln \frac{r}{r_{H}}+2Rl^{2}(r-r_{H})\right]}
{(r_{H}-\frac{q^{2}l^{2}}{r_{H}}+Rl^{2})^{2}}.
 \ee
Here, we have the relation with a constant acceleration $a_{3}$,
 \be
 a_{6}^{2}-a_{3}^{2}=-\frac{1}{l^{2}}
   +\frac{\frac{q^{4}l^{2}}{r^{2}r_{H}^{2}}-\frac{q^{2}}{r_{H}^{2}}g(r)-\frac{2Rq^{2}l^{2}}{rr_{H}(r+r_{H})}}
   {1-\frac{q^{2}l^{2}}{r_{H}^{2}}g(r)+\frac{2Rl^{2}}{r+r_{H}}},
 \ee
which becomes Eq. (\ref{a3a20}) in the limit of $R=0$ and Eq.
(\ref{relnucmbtz}) in the limit of $q=0$.

Exploiting the relation in Eq. (\ref{ths0}), we arrive at the
Unruh temperature for a uniformly accelerated observer in the
(3+3)-dimensional flat spacetime:
 \be
 T_U=\frac{r_{H}-\frac{q^{2}l^{2}}{r_{H}}+Rl^{2}}
{2\pi l\left[r^{2}-r_{H}^{2}-2q^{2}l^{2}\ln \frac{r}{r_{H}}+2Rl^{2}(r-r_{H})\right]^{1/2}}.
 \ee
This is exactly the same with the fiducial temperature $T_{\rm
FID}$ in Eq. (\ref{tfid0}) with $T_{H}$ being the Hawking
temperature measured by an asymptotic observer,
 \be
 T_H=\frac{1}{2\pi}\left(\frac{r_{H}}{l^{2}}-\frac{q^{2}}{r_{H}}+R\right).
 \label{hawkingthqr}
 \ee
As a result, one can say that the Hawking effect for a fiducial
observer in the black hole spacetime is equal to the Unruh effect
for a uniformly accelerated observer in a higher-dimensional flat
spacetime.

In terms of the dimensionless variables, the Hawking temperature
in Eq. (\ref{hawkingthqr}) can be rewritten as
 \be
 T_H\cdot r_{H}=\frac{1}{2\pi}\left(\frac{1}{a^{2}}-b+d\right).
 \label{hawkingthqrless}
 \ee

\begin{figure*}[t!]
 \centering
 \includegraphics{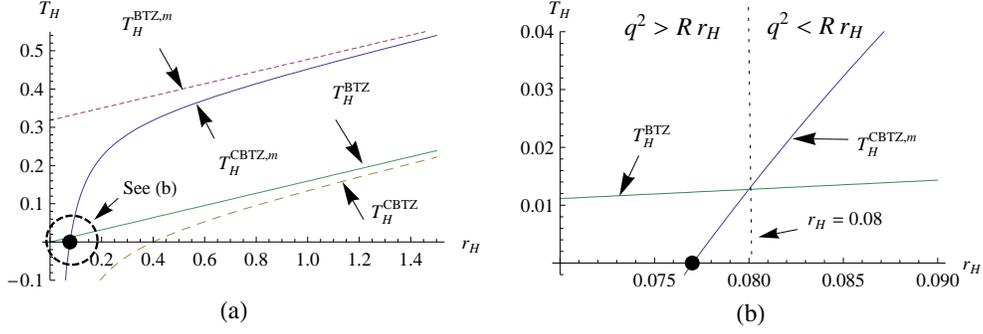}
  \caption{(a) Hawking temperatures for the charged/uncharged BTZ black hole in the massive gravity
   $T^{\rm CBTZ,m}_H$/$T^{\rm BTZ,m}_H$ and
   the charged/uncharged BTZ black hole in the massless gravity $T^{\rm CBTZ}_H$/$T^{\rm BTZ}_H$.
   Here, we have chosen $q=0.4$, $R=2$, and $l=1$ for the charged BTZ black hole
   in the massive gravity. (b) The relative effect of charge and massive
   gravitons in the charged BTZ black hole in the massive gravity:
   $T^{\rm BTZ}_H > T^{\rm BTZ,m}_H$ when $q^2>R r_H$
   and $T^{\rm BTZ}_H < T^{\rm BTZ,m}_H$ when $q^2<R r_H$.  }
 \label{fig4}
\end{figure*}

In Fig. \ref{fig4}(a), we have drawn the Hawking temperatures
$T^{\rm CBTZ,m}_H$/$T^{\rm BTZ,m}_H$ of the charged/uncharged BTZ
black hole in the massive gravity compared to $T^{\rm
CBTZ}_H$/$T^{\rm BTZ}_H$ of the charged/uncharged BTZ black hole
in the massless gravity. One can see that the massive gravitons in
the BTZ black holes only make the Hawking temperatures shift
parallelly. As $r_H\rightarrow\infty$, all the Hawking
temperatures are proportional to $r_H$, while they are being
curved near the event horizon when either $q$ is large or $r_H$ is
small. In Fig. \ref{fig4}(b), one can understand the relative
roles of charge and massive gravitons in the charged BTZ black
hole in the massive gravity that when $q^2>Rr_H$,  $T^{\rm BTZ}_H
> T^{\rm BTZ,m}_H$, while when $q^2<Rr_H$,  $T^{\rm BTZ}_H <
T^{\rm BTZ,m}_H$.

\subsection{Free-fall temperature of charged BTZ black hole in massive gravity}

Now, we assume that an observer at rest is freely falling from the
radial position $r=r_{0}$ at
$\tau=0$~\cite{Brynjolfsson:2008uc,Kim:2009ha,Kim:2013wpa,Kim:2015wwa,hong2015}.
The equations of motion for the orbit of the observer are given by
Eq. (\ref{eomr0}). Exploiting Eqs. (\ref{gemsbtz4}) and
(\ref{eomr0}), we obtain the freely falling acceleration
$\bar{a}_{6}$ in the (3+3)-dimensional GEMS embedded spacetime,
 \be
 \bar{a}_{6}=\frac{1}{l}\left[\frac{\left(1+\frac{q^{2}l^{2}}{rr_{H}}\right)
\left(1-\frac{q^{2}l^{2}}{rr_{H}}+\frac{2Rl^{2}}{r+r_{H}}\right)}
{1-\frac{q^{2}l^{2}}{r_{H}^{2}}g(r)+\frac{2Rl^{2}}{r+r_{H}}}\right]^{1/2},\label{a4btz4}
 \ee
which gives us the temperature measured by the freely falling
observer at rest,
 \be
  T_{\rm FFAR}=\frac{1}{2\pi l}\left[\frac{\left(1+\frac{q^{2}l^{2}}{rr_{H}}\right)
\left(1-\frac{q^{2}l^{2}}{rr_{H}}+\frac{2Rl^{2}}{r+r_{H}}\right)}
{1-\frac{q^{2}l^{2}}{r_{H}^{2}}g(r)+\frac{2Rl^{2}}{r+r_{H}}}\right]^{1/2}.
 \label{tffarcbtzm}
 \ee
\begin{figure*}[t!]
 \centering
 \includegraphics{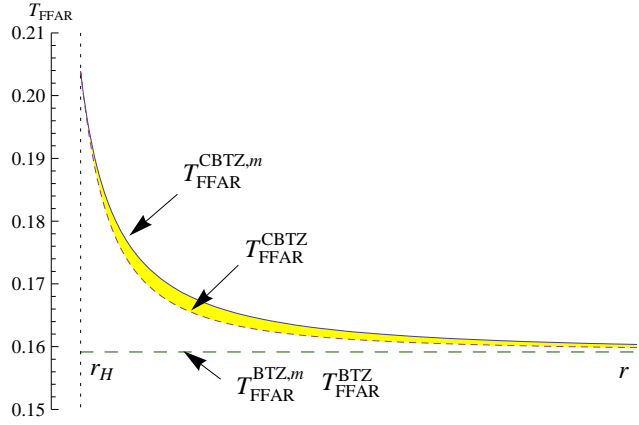}
 \caption{Free-fall temperatures for the charged/uncharged BTZ black hole in
 the massive/massless gravity. The gap between the charged BTZ black holes
 in the massive/massless gravity comes from the gravitons in the massive gravity.
 The free-fall temperatures for the uncharged BTZ black holes in the massive/massless gravity
 are constant all over $r\ge r_H$. Here, we have chosen $r_H=0.5$, $q=0.4$, $R=2$, and $l=1$
 for the charged BTZ black hole in the massive gravity.}
 \label{fig5}
\end{figure*}

It is appropriate to comment that in the limit of $R=0$ the
free-fall temperature in Eq. (\ref{tffarcbtzm}) is reduced to Eq.
(\ref{tffarbtzdef0massless}), as expected. Similarly, one can
readily check that in the $R=0$ limit the physical results
possessing $R$ terms obtained in this section are reduced to those
for the charged BTZ black hole in the massless gravity. On the
other hand, in the $q=0$ limit, the physical results possessing
$q$ terms obtained in this section are also reduced to those for
the uncharged BTZ black hole in the massive gravity discussed in
the previous section.

\begin{figure*}[t!]
 \centering
 \includegraphics{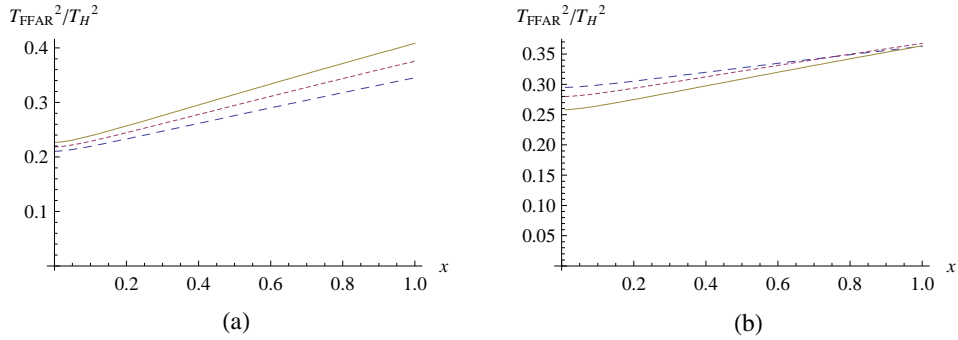}
 \caption{Free-fall temperature for the charged BTZ black hole in
 the massive gravity: (a) for a fixed $d=1$ with $a=2$ and
 $b=0.16,~0.18,~0.20$  from bottom to top; (b) for a fixed $d=1$
 with $b=0.16$ and $a=1.2,~1.4,~1.6$ from top to bottom.}
 \label{fig6}
\end{figure*}
\begin{figure*}[t!]
   \centering
   \includegraphics[scale=0.8]{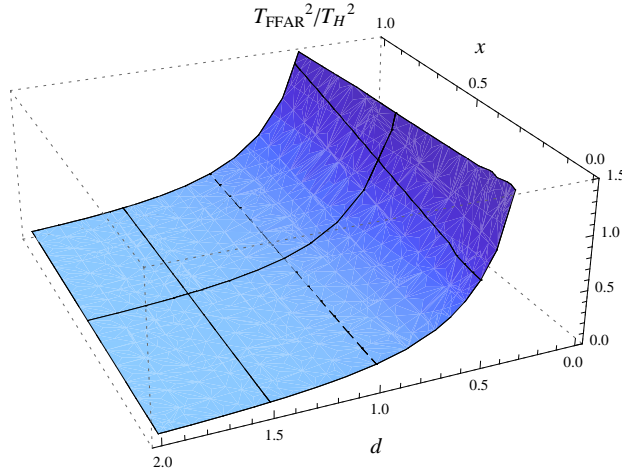}
\caption{Free-fall temperature for the charged BTZ black hole in
the massive gravity with varying $d$ for a fixed $a=2$ and
$b=0.18$. The dashed line represents the curve of $d=1$ with $a=2$
and $b=0.18$ in Fig. \ref{fig6}(a).}
 \label{fig7}
\end{figure*}

In Fig. \ref{fig5}, we have drawn the free-fall temperatures for
the charged/uncharged BTZ black hole in the massive/massless
gravity. Here, one can see the gap between the charged BTZ black
holes in the massive/massless gravity, while they are the same at
both $r_H$ and $r\rightarrow\infty$. This gap arises from the
massive gravitons of the charged BTZ black holes in the massive
gravity. On the other hand, the free-fall temperatures of the
uncharged BTZ black holes in the massive/massless gravity remain
constant all over $r\ge r_H$. This means that freely falling
observers feel the temperature insensitive to the mass term if
they freely fall in an uncharged BTZ black hole.

Now, introducing the dimensionless variables in Eqs.
(\ref{dimensionless0}) and (\ref{dimensionless}), we rewrite the
free-fall temperature in Eq. (\ref{tffarcbtzm}) as
 \be T_{\rm FFAR}\cdot r_{H}=\frac{1}{2\pi a}
    \left[\frac{(1+a^{2}bx)\left(1-a^{2}bx+\frac{2a^{2}dx}{1+x}\right)}
  {1+\frac{2a^{2}bx^{2}}{1-x^{2}}\ln x+\frac{2a^{2}dx}{1+x}}\right]^{1/2}.
  \label{tffdimensionless4}
 \ee
Here, the relevant range of $x$ is given by $0\le x\le 1$. In the
limit of $x=0$, we obtain
 \be
 T_{\rm FFAR}\cdot r_{H}~(x=0)=\frac{1}{2\pi a},
 \label{tffarx04}
 \ee
as in Eq. (\ref{tffarx0massless}). Here, we note that the above
value in Eq. (\ref{tffarx04}) does not depend on the parameters
$b$ and $d$. This means that, as $r\rightarrow\infty$, the charge
and mass term of the charged BTZ black hole in the massive gravity
do not contribute to the free-fall temperature of $T_{\rm FFAR}$
as shown in Fig. \ref{fig5}. Moreover, all the free-fall
temperature $T_{\rm FFAR}$ as $r\rightarrow\infty$ is the same as
that of the uncharged BTZ black hole in the massless gravity in
Eq. (\ref{uchargedtff}).

On the other hand, at $x=1$ corresponding to $r=r_{H}$, we are
left with \be T_{\rm FFAR}\cdot r_{H}~(x=1)=\frac{1}{2\pi
a}(1+a^{2}b)^{1/2}. \label{tffarrh4} \ee This free-fall
temperature in Eq. (\ref{tffarrh4}) is the same as that of the
charged BTZ black hole in the massless gravity in Eq.
(\ref{tffarrhmassless}). Here, we note that at event horizon
$r=r_{H}$ there exists no dependence on the parameter $d$ in the
above free-fall temperature in Eq. (\ref{tffarrh4}), even though
we consider the massive gravity effect of the charged BTZ black
hole case.

We have depicted in Fig. \ref{fig6} the free-fall temperature
$T^2_{\rm FFAR}$ in units of $T^2_H$ for a fixed $d=1$. Figure
\ref{fig6}(a) was drawn by changing the charge $b$, while Fig.
\ref{fig6}(b) was drawn by changing the cosmological constant. The
free-fall temperature for varying $d$ is depicted in Fig.
\ref{fig7}, where it shows that if $d$ is small the whole
variation of $T^2_{\rm FFAR}$ in $T^2_H$ is large in the range of
$0\le x\le 1$.

\section{Discussion}

In summary, we have globally embedded a charged BTZ black hole in
the massive/massless gravity into a (3+3)-dimensional flat
spacetime, while having embedded an uncharged BTZ black hole in
the massless/massive gravity into a (2+2)/(3+2)-dimensional flat
spacetime as shown in the Table \ref{table:table1}.

 \begin{table}[ht]
 \caption{Various GEMS embedding dimensions}
 \centering
 \begin{tabular}{|c|c|c|c|c|}
 \hline
 \hline
 ~~ $q$ ~~  & ~~ $R$ ~~  & ~~ Black holes ~&~ Embedding dimensions\\
 \hline
 ~~ $q=0$  ~~  & ~~ $R=0$ ~~  & ~~ Uncharged BTZ  black hole in massless gravity ~&~ (2+2)\\
 \hline
 ~~  $q\neq 0$  ~~  & ~~ $R=0$ ~~  & ~~ Charged BTZ black hole in massless gravity ~&~ (3+3)\\
 \hline
 ~~  $q=0$  ~~  & ~~ $R\neq 0$ ~~  & ~~ Uncharged BTZ black hole in massive gravity ~&~ (3+2)\\
 \hline
 ~~  $q\neq 0$  ~~  & ~~ $R\neq 0$ ~~  & ~~ Charged BTZ black hole in massive gravity ~&~ (3+3)\\
 \hline
 \end{tabular}
 \label{table:table1}
 \end{table}

Making use of the embedded coordinates, we have directly obtained
the Unruh, Hawking and freely falling temperatures in the
(un)charged BTZ black hole in the massive/massless gravity and
shown that the Hawking effect for a fiducial observer in a curved
spacetime is equal to the Unruh effect for a uniformly accelerated
observer in a higher-dimensionally embedded flat spacetime.
Moreover, we have evaluated all the free-fall temperatures of the
(un)charged BTZ black hole in the massive/massless gravity
measured by freely falling observers into a black hole. As a
result, we have found that all the free-fall temperatures given by
$\frac{1}{2\pi l}$ at the infinity end up hotter but finite at the
horizon when black holes are charged, while remaining the same as
they start when black holes are uncharged like in Table
\ref{table:table2}.

{\renewcommand{\arraystretch}{2}%
 \begin{table}[ht]
 \caption{Free-fall temperature at the infinity and event horizon}
 \centering
 \begin{tabular}{|c|c|c|}
 \hline
 \hline
 ~ BTZ black holes ~  & ~ $r\rightarrow\infty$ ~  & ~$r\rightarrow r_H$ \\
 \hline
 ~ uncharged BTZ black hole in massless/massive gravity ~  & ~$\frac{1}{2\pi l}$~  & ~$\frac{1}{2\pi l}$ \\
 \hline
  ~charged BTZ black hole in massless/massive gravity ~  & ~$\frac{1}{2\pi l}$~  & ~$\frac{1}{2\pi l}\left(1+\frac{q^2l^2}{r^2_H}\right)^{1/2}$\\
 \hline
  \end{tabular}
 \label{table:table2}
 \end{table}

\acknowledgements{One of us (Y. W. K.) was supported by the
National Research Foundation of Korea grant funded by the Korea
government, NRF-2017R1A2B4011702.}


\end{document}